\documentclass[preprint,aps,amsmath,pre]{revtex4-1}
\usepackage{graphicx}
\usepackage{amsmath}
\usepackage{caption}
\usepackage{subcaption}

\begin{document}

\title{Evolutionary consequences of assortativeness in haploid
genotypes}

\author{David M. Schneider$^1$, Ayana B. Martins$^2$, Eduardo do Carmo$^3$,
and Marcus A.M. de Aguiar$^1$}

\affiliation{$^1$ Instituto de F\'{\i}sica `Gleb Wataghin',
Universidade Estadual de Campinas, Unicamp\\ 13083-859, Campinas, SP,
Brasil\\$^2$Instituto de Bioci\^encias, Universidade de S\~ao Paulo,
05508-090, S\~ao Paulo, SP, Brazil\\$^3$ Universidade Federal da
Integra\c{c}\~ao Latino Americana, 85867-970, Foz do Igua\c{c}u, PR,
Brazil.}

\begin{abstract}

We study the evolution of allele frequencies in a large population where
random mating is violated in a particular way that is related to recent works
on speciation. Specifically, we consider non-random encounters in
haploid organisms described by biallelic genes at two loci and assume that individuals
whose alleles differ at both loci are incompatible. We show that evolution under
these conditions leads to the disappearance of one of the alleles and
substantially reduces the diversity of the population. The allele that 
disappears, and the other allele frequencies at equilibrium, depend only 
on their initial values, and so does the time to equilibration. However, certain
combinations of allele frequencies remain  constant during the process,
revealing the emergence of strong correlation between the two loci promoted by
the epistatic mechanism of incompatibility.  We determine the geometrical 
structure of the haplotype frequency space and solve the dynamical equations, 
obtaining a simple rule to determine equilibrium solution from the initial 
conditions. We show that our results are equivalent to selection against double 
heterozigotes for a population of diploid individuals and discuss the 
relevance of our findings to speciation. 

\end{abstract}

\maketitle

\section{Introduction}

While the origin of species has always been a central subject in evolutionary 
biology, the large number of recent empirical and theoretical 
developments has renewed the interest in the area 
\citep{coyne_speciation_2004,gavrilets_2004,butlin_what_2012,
nosil_ecological_2012}.  Individual-based simulations, in particular, have been
successful in fostering  relevant discussions in speciation
\citep{higgs1991,higgs1992,kondra1998,dieckmann_origin_1999,
gavrilets_dynamical_1999,doursat2008,doorn_origin_2009,fitzpatrick2009,
de_aguiar_global_2009,ashlock2010,kopp2010,melian2011,desjardins2012}.
Specifically, simulations in which mating is restricted by spatial and genetic
distances have been able to describe empirical patterns of species diversity
\citep{de_aguiar_global_2009} and within-species genetical diversity
\citep{martins_evolution_2013}.

One of the simplest ways of introducing assortativeness in mating in a 
individual-based simulation is to attribute haploid genomes with $B$ 
biallelic loci to individuals and allow them to mate only if the genomes 
differ in no more than $G$ loci
\citep{gavrilets_patterns_2000,de_aguiar_global_2009,martins_evolution_2013}.
This approach considers that mate choice often relies on multiple cues that are
determined genetically \citep{candolin_use_2003}. In the case
of assortative mating, we assume that individuals have a certain tolerance to
differences when choosing a mate, however if the other individual is too
different, it will no longer be considered a potential mate. Under these
assumptions, reproductive isolation was shown to be maintained among demes in
the presence of sufficiently low migration rates
\citep{gavrilets_patterns_2000}. Spatially explicit versions of this
process have also been studied
\cite{doursat2008,fitzpatrick2009,de_aguiar_global_2009,ashlock2010,melian2011,
martins_evolution_2013} and, in particular, speciation was shown to
emerge spontaneously if mating is also constrained by the spatial
distance \citep{doursat2008,de_aguiar_global_2009,ashlock2010,
martins_evolution_2013}.

In order to reflect the dynamics of evolving populations, most simulations
need to incorporate several ingredients simultaneously, such as mutation, 
genetic drift, recombination,  assortativeness in mating and individual's 
movement and spatial positioning. Gavrilets \cite{gavrilets_dynamical_1999}
proposed and analysed a number of simplified mathematical models that are
closely related to these simulations, including selection, mutation, drift and
population structure. These more realistic approaches to speciation do not 
allow for the detailed understanding of how each of the mechanisms involved 
contribute to the emergence and maintenance of reproductive isolation.

To construct a dynamical theory that accounts for the predictions of the
model described in \cite{de_aguiar_global_2009} and other similar models, it is
important to understand the roles of their different ingredients and to validate
their generality. It has already been shown that separation of individuals into
males and females does not introduce important effects in the conditions for speciation
\cite{baptestini_2013,schneider_2012}, originally based on hermaphrodite
populations. In this paper we focus on the effect of genetic incompatibility and
work out the theory for infinitely large populations with two biallelic loci
($B=2$) without mutations. Genetic incompatibilities will be implemented by
allowing reproduction only if the alleles from the parents differ at most in
one locus ($G=1$). This is the simplest system for which the genetical mechanism
of interest may be implemented. We will show that this process leads to
evolution by changing the allele frequencies and that it is one of the main
ingredients in the process of speciation studied in
\citep{de_aguiar_global_2009}. Despite the changes in all 
allele frequencies, we will demonstrate that a certain combination of
frequencies from the two loci remain constant during the evolution, revealing a
strong correlation between the loci introduced by the genetic mating restriction. 

The paper is organized as follows. In section \ref{mod_reprod} we describe the
reproductive mechanism employed in the dynamics. In section \ref{no_restrictions}
we characterize the evolution of a population subjected to no mating restrictions 
(random mating), which is similar to the Hardy-Weinberg (HW) equilibrium.
The mathematical implications of the genetic restriction, including the
description of equilibria and their features, are  analyzed in section  
\ref{con_restr}. Finally, in section \ref{conclusions} we expose our conclusions
and discuss the  possible evolutionary impacts of our results. Mathematical
technicalities not strictly essential to the discussion are included in the
appendices.

\section{Reproductive mechanism}\label{mod_reprod}

Consider a population of $N$  hermaphrodite individuals with haplotypes $AB$,
$Ab$, $aB$, and $ab$ ($A$ and $a$ being the alleles at the locus 1,
and $B$ and $b$ the alleles at the locus 2), whose composition at the generation
$t$ is characterized by the numbers $N_{AB}$, $N_{Ab}$, $N_{aB}$ and $N_{ab}$
($\sum_{u',u''} N_{u'u''}=N$, with $u'=\{A,a\}$ and $u''=\{B,b\}$).
All possibles encounters between members of this generation give 
an offspring which will be a member of the  generation $t+1$ with a probability
$r_{h_1\!:h_2}$, $h_1$ and $h_2$ being the paternal haplotypes (we include in
$r_{h_1\!:h_2}$  both effects of compatibility of  the parents and the viability
of the new born individual). By assuming no overlap among generations, the
contributions to the individuals with  haplotype $AB$ at generation $t+1$  
can be inferred from Table \ref{table1}.
\begin{table}
 \[
\begin{array}{c|c|c}
\text{Paternal haplotypes} & \text{Number of encounters} & \text{Fraction of successful}\\
                           &                             & \text{$AB$ offspring}\\
\hline  
AB \times AB               &     \frac{1}{2}N_{AB}\times(N_{AB}-1)&     r_{AB:AB}\\
AB \times Ab               &          N_{AB}\times N_{Ab}        &        1/2\times r_{AB:Ab}\\
AB \times aB               &          N_{AB}\times N_{aB}        &        1/2\times r_{AB:aB} \\
AB \times ab               &          N_{AB}\times N_{ab}        &        1/4 \times r_{AB:ab}\\
Ab \times aB               &          N_{Ab}\times N_{aB}        &        1/4 \times r_{Ab:aB}\\
\end{array}
\]
\caption{Production of individuals with haplotypes $AB$ at generation $t+1$,
through encounters between individuals of the generation $t$.\label{table1}}
\end{table}

The number  of $AB$ individuals at time $t+1$ obeys thus the equation
\begin{equation}\label{conN_1}
 N_{AB}^{t+1}=\frac{N_{AB}(N_{AB}-1)}{2} r_{AB:AB}+
   \frac{N_{AB} N_{Ab}}{2} r_{AB:Ab}+\frac{N_{AB} N_{aB}}{2} r_{AB:aB}+\frac{N_{AB} N_{ab}}{4} r_{AB:ab}+
   \frac{N_{Ab} N_{aB}}{4} r_{Ab:aB}
\end{equation}

Equivalent tables allow  to obtain evolution equations for  the remaining
haplotypes

\begin{equation}\label{conN_2}
  N_{Ab}^{t+1}=\frac{N_{Ab}(N_{Ab}-1)}{2} r_{Ab:Ab}+
   \frac{ N_{AB} N_{Ab}}{2} r_{AB:Ab}+\frac{N_{Ab} N_{ab}}{2} r_{Ab:ab}+\frac{N_{AB} N_{ab}}{4} r_{AB:ab}+
   \frac{N_{Ab} N_{aB}}{4} r_{Ab:aB}
\end{equation}

\begin{equation}\label{conN_3}
 N_{aB}^{t+1}=\frac{N_{aB}(N_{aB}-1)}{2} r_{aB:aB}+
  \frac{N_{AB} N_{aB}}{2} r_{AB:aB}+\frac{N_{aB} N_{ab}}{2} r_{aB:ab}+\frac{N_{AB} N_{ab}}{4} r_{AB:ab}
  +\frac{N_{Ab} N_{aB}}{4} r_{Ab:aB}
\end{equation}

\begin{equation}\label{conN_4}
 N_{ab}^{t+1}=\frac{N_{ab}(N_{ab}-1)}{2} r_{ab:ab}+
   \frac{N_{Ab} N_{ab}}{2} r_{Ab:ab}+\frac{N_{aB} N_{ab}}{2} r_{aB:ab}+\frac{N_{AB} N_{ab}}{4} r_{AB:ab}
  +\frac{N_{Ab} N_{aB}}{4} r_{Ab:aB}
 \end{equation}

In the following sections we analyze the dynamics of the haplotype frequencies
$p_{u'u''}\equiv N_{u'u''}/N$ in the infinite limit of the population size. For
each different scenario we specify the values of the probabilities
$r_{h_1\!:h_2}$ by setting the total number of individuals constant along
generations.
  
\section{The non restricted case}\label{no_restrictions}

This section summarizes the outcomes for the case of no genetic
restrictions. Although some
of the results described in here can be found in the literature (see for example
\cite{crow_introduction_1970,ewens}), the following  discussion is fundamental
as a reference for comparing the results presented next.

If random mating is assumed, $r_{h_1\!:h_2}=r$ for every encounter. Substituting in 
equations (\ref{conN_1}-\ref{conN_4})  and summing up, one  obtains

\begin{equation}
N^{t+1}= r N(N-1)/2 \equiv N, 
\end{equation}
so that $r=2/N$ for very large populations.
By introducing $D= p_{AB}p_{ab}-p_{Ab}p_{aB}$, the so called linkage 
disequilibrium, and after some algebra, equations for the haplotype frequencies read
\begin{eqnarray}
&p_{AB}^{t+1}&=p_{AB}-\frac{1}{2}D\label{sr11}\\
&p_{Ab}^{t+1}&=p_{Ab}+\frac{1}{2}D\label{sr12}\\
&p_{aB}^{t+1}&=p_{aB}+\frac{1}{2}D\label{sr13}\\
&p_{ab}^{t+1}&=p_{ab}-\frac{1}{2}D\label{sr14}\\\nonumber,
\end{eqnarray}
from which one immediately sees that a sufficient  condition for the equilibrium is $D=0$, or $p_{AB}p_{ab}=p_{Ab}p_{aB}$. 
Notice that the quantities
\begin{eqnarray}
&\tilde{p}_A=&p_{AB}+p_{Ab}\label{allele_A}\\
&\tilde{p}_a=&p_{aB}+p_{ab}\label{allele_a}\\
&\tilde{p}_B=&p_{AB}+p_{aB}\label{allele_B}\\ 
&\tilde{p}_b=&p_{Ab}+p_{ab}\label{allele_b},\\
\nonumber\end{eqnarray}
representing the frequencies of the four available alleles, remain constant
from the first generation. This is also the case in the HW equilibrium context,
however it should be emphasized that in the present framework there are two
independent allele frequencies (because
$\tilde{p}_A+\tilde{p}_a=\tilde{p}_B+\tilde{p}_b=1$) in contrast to the  HW
equilibrium  where the only independent variable is the frequency of one of the
two available alleles.
  
The time dependence of the haplotype frequencies can be obtained analytically
(see appendix \ref{appendixA}). Here we just look for a relationship between the
haplotype  and the allele frequencies.We start calculating  $D$ at time $t+1$,
 \begin{equation}
  D^{t+1}=(p_{AB}-\frac{1}{2}D)(p_{ab}- \frac{1}{2}D)-(p_{Ab}+ \frac{1}{2}D)(p_{aB}+\frac{1}{2}D)=\frac{1}{2}D,
 \end{equation}
whose solution is simply
  \begin{equation}\label{D_de_t}
   D=\frac{1}{2^t}D^0,
  \end{equation}
 $D^0$ being the initial value of $D$.
 Combining (\ref{allele_A}-\ref{allele_B}) and
  (\ref{sr11}-\ref{sr14}), it is possible to deduce the following relationships (see appendix \ref{appendixB})
  \begin{eqnarray}
   &p_{AB}&=\tilde{p}_A \tilde{p}_B+  D=\tilde{p}_A \tilde{p}_B+ \frac{1}{2^t}D^0\label{pAB_t}\\
   &p_{Ab}&=\tilde{p}_A \tilde{p}_b-  D=\tilde{p}_A \tilde{p}_b- \frac{1}{2^t} D^0\label{pAb_t}\\
   &p_{aB}&=\tilde{p}_a \tilde{p}_B-  D=\tilde{p}_a \tilde{p}_B- \frac{1}{2^t} D^0\label{paB_t}\\
   &p_{ab}&=\tilde{p}_a \tilde{p}_b+  D=\tilde{p}_a \tilde{p}_b+ \frac{1}{2^t} D^0\label{pab_t}\\
   \nonumber
 \end{eqnarray}

Accordingly, the haplotype frequencies reach an equilibrium asymptotically and, 
as in the case of the HW equilibrium, is related to
the constant alleles frequencies,
 \begin{eqnarray}
   &p_{AB}^{\rm eq}&=\tilde{p}_A \tilde{p}_B\label{constant_allele_frequencies_1}\\
   &p_{Ab}^{\rm eq}&=\tilde{p}_A \tilde{p}_b\label{constant_allele_frequencies_2}\\
   &p_{aB}^{\rm eq}&=\tilde{p}_a \tilde{p}_B\label{constant_allele_frequencies_3}\\
   &p_{ab}^{\rm eq}&=\tilde{p}_a \tilde{p}_b\label{constant_allele_frequencies_4}\\
   \nonumber
 \end{eqnarray}
It is important to remark the asymptotic  behavior of the haplotype frequencies
toward the equilibrium (equations (\ref{pAB_t}-\ref{pab_t})), in contrast to  HW
theorem in which the equilibrium of the genotype frequencies is attained in one
generation.

%%%%%%%%%%%%%%%%%%%%%%%%%%%%%%%%%%%%%%%%%%%%%%%%%%%%%%%%%%%%%%%%%%%%%%%%%%%%%%%%%%%%%%%%%%%%%%%%%%%%%
 \section{Genetically restricted mating}\label{con_restr}

To mathematically describe the mating restriction imposed to individuals
differing in more than one allele, we simply redefine the
compatibility-viability rate as follows 

\begin{equation}\label{rh1h2}
 r_{h_1:h_2}=
 \left\{
 \begin{array}{cc}
  0 & h_1\!:\!h_2=AB\!:\!ab \;\text{ or }\; h_1\!:\!h_2=Ab\!:\!aB \\
  r' & {\rm otherwise}
 \end{array}
\right.
\end{equation}

Following the procedure of section \ref{no_restrictions}, we obtain
\begin{equation}\label{rPrime}
 r'=\frac{2}{N}\frac{1}{1-2\Delta}
\end{equation}
with 
\begin{equation}
\Delta\equiv \frac{N_{AB}N_{ab}+N_{Ab}N_{aB}}{N^2}=p_{AB}p_{ab}+p_{Ab}p_{aB}
\end{equation}

Notice that $r'$ is not constant, in contrast to the rate $r$ of section 
\ref{no_restrictions},  but varies along generations  depending on how many incompatible 
encounters may take place. The more incompatible encounters, the bigger the chance
of a compatible encounter to give an offspring viable for the next generation.

By substituting (\ref{rPrime}) in (\ref{rh1h2}), and (\ref{rh1h2}) in 
(\ref{conN_1}-\ref{conN_4}), equations for the haplotype frequencies reduce to 
 
\begin{eqnarray}
p_{AB}^{t+1}= \frac{p_{AB}(1-p_{ab})}{1-2\Delta}\label{freq_comRes_1}\\
p_{Ab}^{t+1}= \frac{p_{Ab}(1-p_{aB})}{1-2\Delta}\label{freq_comRes_2}\\
p_{aB}^{t+1}= \frac{p_{aB}(1-p_{Ab})}{1-2\Delta}\label{freq_comRes_3}\\
p_{ab}^{t+1}= \frac{p_{ab}(1-p_{AB})}{1-2\Delta}\label{freq_comRes_4}\\
\nonumber
\end{eqnarray}
In what follows, we explore the dynamics governed by equations
(\ref{freq_comRes_1}-\ref{freq_comRes_4}) on the basis of a stability
analysis.

%%%%%%%%%%%%%%%%%%%%%%%%%%%%%%%%%%%%%%%%%%%%%%%%%%%%%%%%%%%%%%%%%%%%%%%%%%%%%%%%%%%%%%%%%%%%%%%%%%%%%
\subsection{Equilibrium solutions and stability analysis}\label{stab_anal}

Equations (\ref{freq_comRes_1}-\ref{freq_comRes_4}) display four different types
of fixed points, summarized in Table \ref{fixedPoints}. As we will see next,
only types 1 and 2 are stable.

\begin{table}
\caption{\label{fixedPoints}The four types of fixed points of the dynamical
system  (\ref{freq_comRes_1}-\ref{freq_comRes_4}). For equilibria of type 1 and
2, the label subscripts indicate the alleles which are lost.}
\begin{tabular}{ l c c c}
\hline \\
Type & Label & Coordinates &Stability\\  \hline \\ 
Type 1. Continuous sets. Two & $E_A$ & 
$p_{AB} = p_{Ab} = 0, p_{aB} = \lambda_A, p_{ab} = 1- \lambda_A$&\\
compatible haplotypes have   & $E_B$ & 
$p_{AB} = p_{aB} = 0, p_{Ab} = \lambda_B, p_{ab} = 1 - \lambda_B$& Stable \\
zero frequency; one allele  is   & $E_a$ & 
$p_{aB} = p_{ab} = 0, p_{AB} = \lambda_a, p_{Ab} = 1 - \lambda_a$&\\
 lost in one locus. The other& $E_b$ & 
$p_{Ab} = p_{ab} = 0, p_{AB} = \lambda_b, p_{aB} = 1 -\lambda_b$&\\
  locus remains polymorphic. & &$\lambda_{A,B,a,b} \in (0, 1)$ &\\
\hline \\
Type 2. Three haplotypes have  & $E_{AB}$ & $p_{AB} = p_{Ab} = p_{aB} = 0, p_{ab} = 1$&\\
zero frequency. One allele is lost  & $E_{aB}$ & $p_{AB} = p_{aB} = p_{ab} = 0, p_{Ab} = 1$&Stable\\
 at both loci.   & $E_{ab}$ & $p_{Ab} = p_{aB} = p_{ab} = 0, p_{AB} = 1$ &\\
.& $E_{Ab}$ & $p_{AB} = p_{Ab} = p_{ab} = 0, p_{aB} = 1$&\\ 
\hline \\
Type 3. Two incompatible  & $EU_1$ &  $p_{Ab} = p_{aB} = 0, p_{AB} = p_{ab} = 1/2$&Unstable\\
haplotypes have zero frequency. & $EU_2$ & $p_{AB} = p_{ab} = 0, p_{Ab} = p_{aB} = 1/2$&\\ 
\hline \\
Type 4. Equiprobable  & $ES$ & $p_{AB} = p_{Ab} = p_{aB} = p_{ab} = 1/4$&Saddle\\ 
distribution. & & &\\ 
\hline
\end{tabular}
\end{table}

\begin{figure} \centering
\includegraphics[scale=0.4]{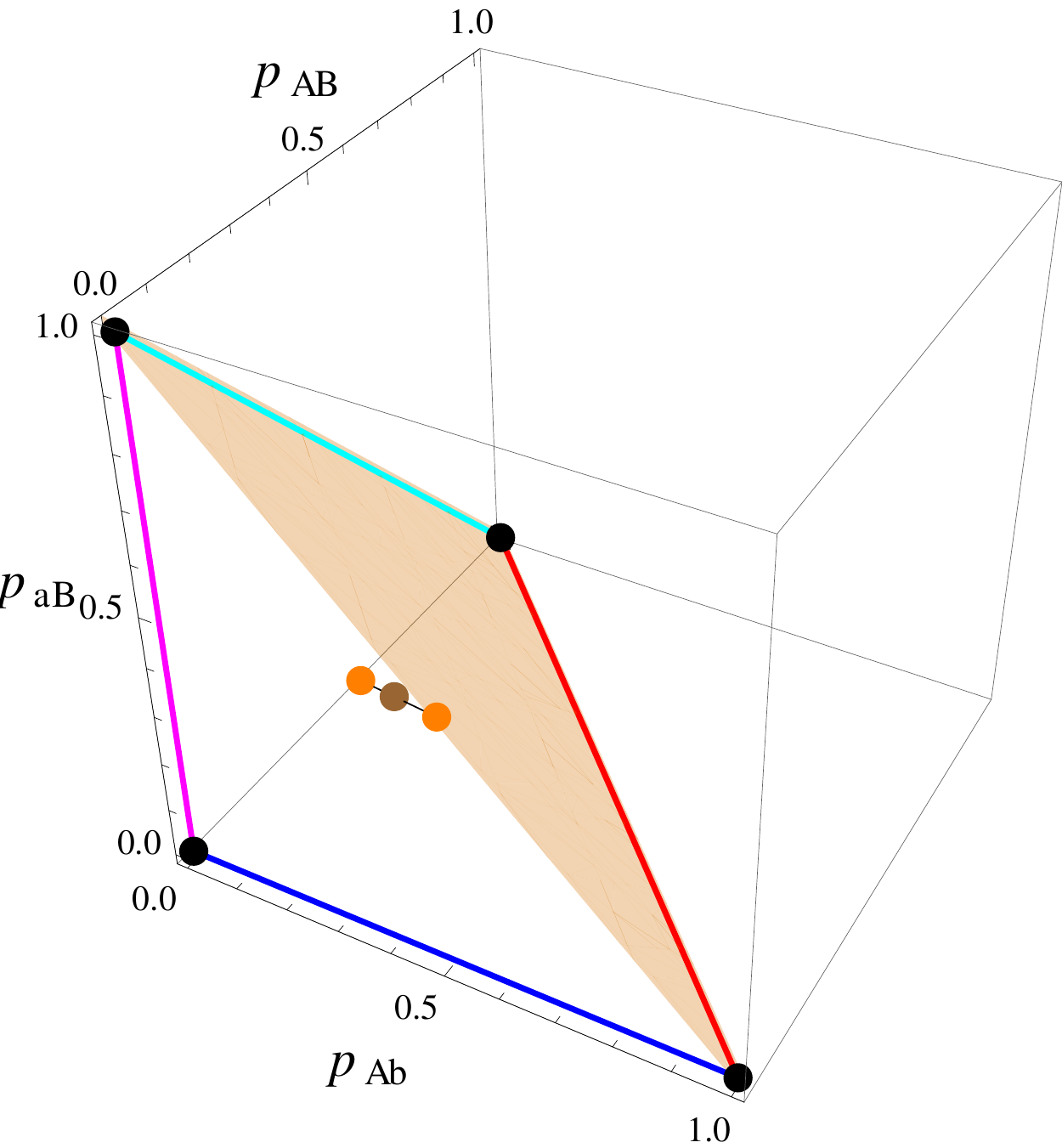}\hfil
\includegraphics[scale=0.3]{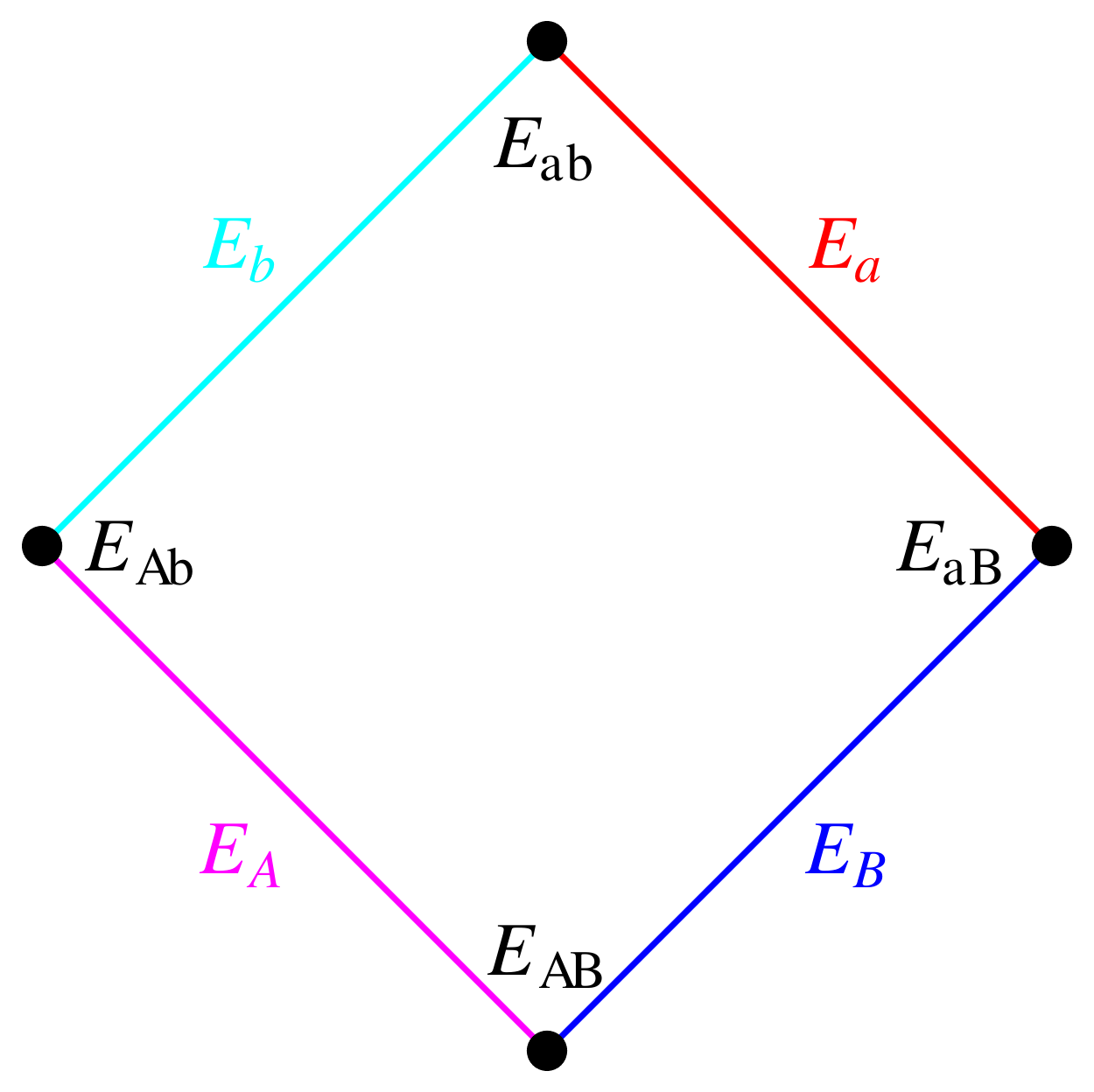}\hfil
\includegraphics[scale=0.4]{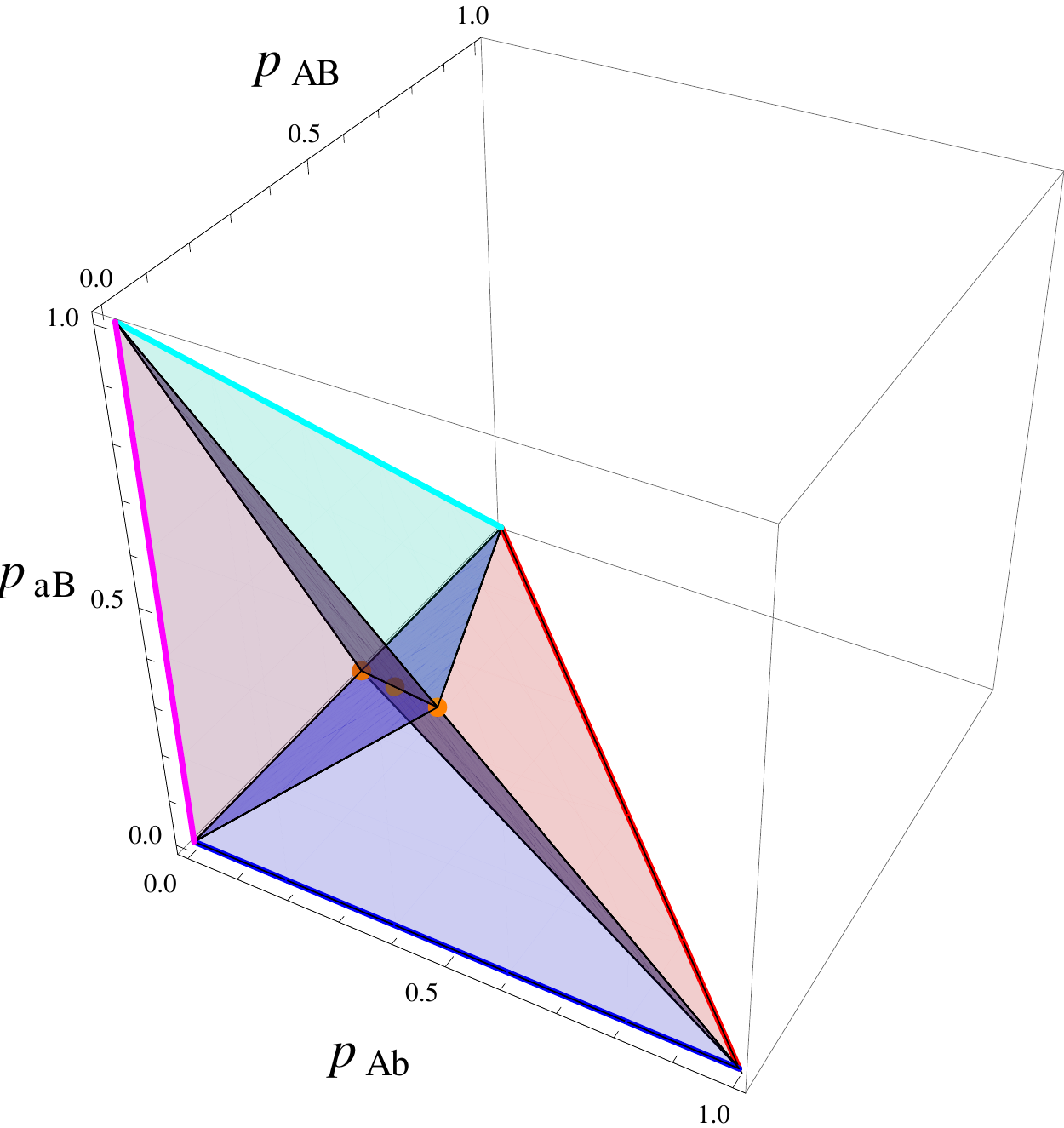}\hfil
\caption{(Color online) Left: 3-dimensional phase space displaying the four
families of equilibrium solutions: $E_A$
($p_{aB}$ axis, purple), $E_B$ ($p_{Ab}$ axis, blue), $E_a$ (diagonal on
the $p_{Ab}-p_{AB}$ plane, red) and $E_b$ (diagonal on the $p_{aB}-p_{AB}$
plane, cyan); $E_{ab}$, $E_{Ab}$, $E_{aB}$ and $E_{AB}$ (vertices connecting
the first family, black circles); $EU_1$ and $EU_2$ (midpoints of edges of the
phase space not  containing the first family, orange circles); and $ES$ (center,
brown circle). The shaded light brown surface represents the top face of the
tetrahedral phase space of equation $p_{AB}+p_{Ab}+p_{aB}= 1$.  Middle:
Schematic representation  of the stable fixed points. Right: Division of the
phase space displaying the basins of attraction of type 1 fixed points.
\label{ptos_fixos}}
\end{figure}

Since $p_{AB}+p_{Ab}+p_{aB}+p_{ab}=1$, it is possible to give a graphical
description of the dynamics by constructing a 3-dimensional phase space.
We arbitrarily chose the frequencies $p_{AB}$, $p_{Ab}$ and $p_{aB}$ as the
independent dynamic variables. The constrains $p_{AB}\ge 0$, $p_{Ab}\ge 0$,
$p_{aB} \ge 0$, and $p_{AB}+p_{Ab}+p_{aB} \le 1$ give the phase space the
geometry of a tetrahedron having right triangular faces (Figure
\ref{ptos_fixos}). The top face of the tetrahedron, defined by the equation
$p_{AB}+p_{Ab}+p_{aB}= 1$, corresponds to frequencies distributions having
$p_{ab}=0$.  $p_{ab}=1$ implies $p_{AB}+p_{Ab}+p_{aB}= 0$ and is represented 
by the origin. Type 1 fixed points  are located at four of the six edges of the
tetrahedron (colored edges in Figure \ref{ptos_fixos}), the points of type 2 
are the vertices of the tetrahedron (black circles), type 3 fixed points $EU_1$
and $EU_2$ are located at the midpoints of the edges not containing points of
type 1 (orange circles) and finally, the center of the tetrahedron houses the 
type 4 fixed point $ES$ (brown circle). 
  
We start the discussion with type 3 fixed points for which the stability
matrix is two times the identity. Therefore, it has one single fully  degenerated 
eigenvalue  $\zeta=2$ and both points $EU_1$ and $EU_2$ are unstable fixed
points.

The stability matrix of $ES$ displays two different eigenvalues, $\zeta_s=2/3$
and $\zeta_u=4/3$, the latter with degeneration 2. Accordingly, this fixed point
has a saddle like behavior, being unstable on a two dimensional subspace and
stable on a one dimensional subspace. From a geometrical point of view, it is
interesting to note that the points $EU_1$ and  $EU_2$ are equidistantly 
located from $ES$ along the linear subspace spanned  by the  stable eigenvector 
${\bf e}_{AB}-{\bf e}_{Ab}-{\bf e}_{aB}$  (Figure \ref{ptos_fixos}). 

Fixed points of types 1 and 2 deserve a more detailed description. Not 
displaying exactly the same properties, they share common features,
which makes  instructive to analyze the stability of both types at the same
time. We take as an example the set of points $E_B$, and its
 $\lambda_B\rightarrow 0$ and $\lambda_B\rightarrow 1$ limits, which are the
points $E_{AB}$ and $E_{aB}$, respectively. Appendix  \ref{appendixC} explains
how to transfer  the outcomes of the following analysis to the remaining type 1
and type 2 fixed points. The stability matrix  for any of such points has the
following eigenvalues and eigenvectors:
 
 \begin{itemize}
 \item  $\zeta_1=\lambda_B$:   ${\mathbf v_1}= \frac{1}{2}{\bf e}_{AB}-\lambda_B {\bf e}_{Ab}$.
 \item  $\zeta_2=1-\lambda_B$: ${\mathbf v_2}=(\frac{1}{2}-\lambda_B){\bf e}_{Ab}+\frac{1}{2}{\bf e}_{aB}$.
 \item $\zeta_3=1$: ${\mathbf v_3}={\bf e}_{Ab}$.
 \end{itemize}
 
In the first place, notice that along the direction spanned by ${\mathbf v_3}$ 
displacements are neutral.  Indeed, since $\zeta_3=1$ for all points in the set,
displacements from the fixed points in this direction are not amplified nor
contracted. This is consistent with the fact that this direction corresponds to
the  $p_{Ab}$ axis itself, where the entire set  $E_B$ is located. Therefore,
by displacing a point from a fixed point in this direction one simply moves to
another fixed point and thus iterations do not evolve it further.

\begin{figure}
\centering
\includegraphics[scale=0.6]{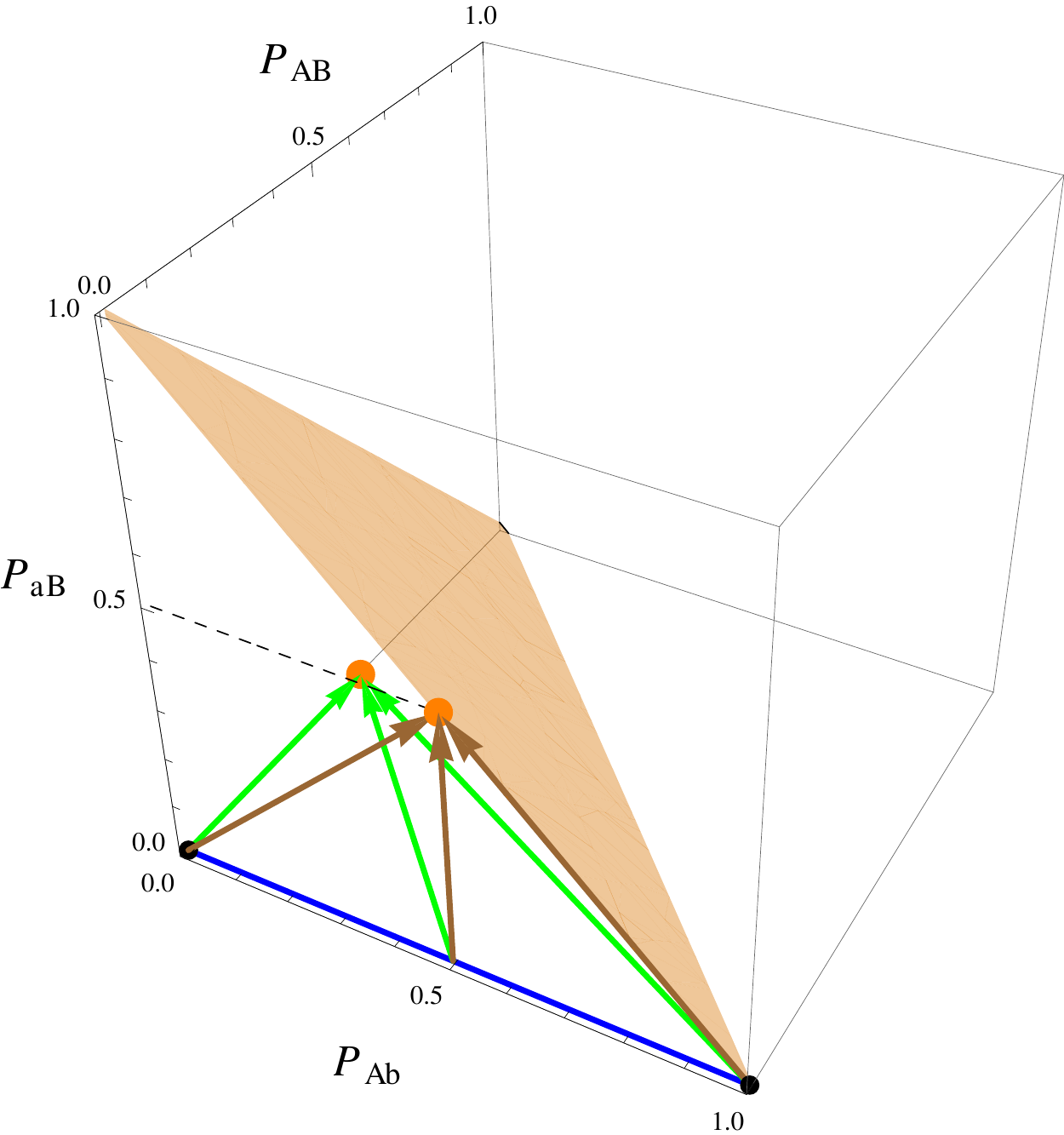}\hfil
\caption{(Color online) Eigenvectors $\mathbf{v_1}$ (green) and $\mathbf{v_2}$
(brown) corresponding to the points $E_{B}$  (with $\lambda_B=1/2$), $E_{AB}$
and $E_{aB}$. }
\label{vectorsitos}	
\end{figure}

In the directions spanned by ${\mathbf v_1}$ and ${\mathbf v_2}$, the
eigenvalues show that the fixed points $E_B$ are stable (points
$E_{AB}$ and $E_{aB}$ are also stable, however the stability can not  be
inferred from the eigenvalues). Notice that, properly scaled, eigenvectors
${\mathbf v_1}$ and ${\mathbf v_2}$ have the interesting property of connecting
the fixed points $E_B$ (as well as $E_{AB}$ and $E_{aB}$) with the points $EU_1$ and
$EU_2$, respectively. This property, illustrated in Figure \ref{vectorsitos},
will be used in section \ref{conserved}.

%%%%%%%%%%%%%%%%%%%%%%%%%%%%%%%%%%%%%%%%%%%%%%%%%%%%%%%%%%%%% 
\subsection{Rates of convergence}
\label{rates}

Although the qualitative behavior of any fixed point in the set $E_B$ 
is the same (one neutral direction and two stable directions
pointing to type 3 fixed points), and even for the extremes 
 $E_{AB}$ and $E_{aB}$, points within the set differ from each other in  the
time to convergence.  Close to the fixed points the movement along a given
eigendirection obeys
\begin{equation}
 x^{t+1}_i=\zeta_i x_i,
 \end{equation} 
so that 
\begin{equation}\label{soluc_p1}
x_i= x_i^0 \zeta_i^t \equiv x_i^0 e^{-t/\tau_i},
\end{equation}
($x_i^0$ being the i-th component of the initial displacement from the fixed
point, for $i=1,2$) with	
\begin{equation}\label{time_estimation}
  \tau_i = -\frac{1}{\ln{\zeta_i}}.
\end{equation}
This allows to compare the time constants in the  directions ${\mathbf v_1}$ 
and ${\mathbf v_2}$ as the parameter $\lambda_B$ varies along the set.
The ratio gives
\begin{equation}
\frac{\tau_1}{\tau_2}= 
\frac{\ln{(1-\lambda_B)}}{\ln{\lambda_B}}.
\end{equation}
Accordingly, by displacing the fixed point close to the point $E_{AB}$ 
($\lambda_B \sim 0$), the time  of convergence along ${\mathbf v_2}$ becomes
much larger compared to the time along ${\mathbf v_1}$. The opposite behavior is
obtained  by displacing the fixed point  towards $E_{aB}$ ($\lambda_B\sim1$). 

For $\lambda_B$ strictly equal to zero,  estimation (\ref{time_estimation})
yields an infinitely slow convergence  along  ${\mathbf v_2}$, and an instantaneous 
convergence along ${\mathbf v_1}$. This is however a consequence of
the attempt to linearize an equation with no linear contribution in its series
expansion. Since $p_{Ab}=p_{aB} =0$ along ${\mathbf v_1}$, we can rewrite
equation (\ref{freq_comRes_1})  as
\begin{equation}
 p_{AB}^{t+1}=\frac{p_{AB}^2}{1-2p_{AB}(1-p_{AB})},
\end{equation}
whose leading order is quadratic. We write therefore 
\begin{equation}
 p_{AB}^{t+1}=p_{AB}^2+O(p_{AB}^3)
\end{equation}
for points close to $E_{AB}$, whose leading order solution reads
\begin{equation}
p_{AB}= (p_{AB}^0)^{2^t}.
\end{equation}
Besides demonstrating stability, this solution shows that convergence is
superfast in comparison to the exponential behavior for  points $E_{B}$
(equation (\ref{soluc_p1})).

In the ${\mathbf v_2}$ direction $p_{Ab}=p_{aB}$ and $p_{AB}=0$. Therefore, we 
rewrite equation (\ref{freq_comRes_2}) as
\begin{equation}\label{P01paralambda0v2}
 p_{Ab}^{t+1}=\frac{p_{Ab}(1-p_{Ab})}{1-2
 p_{Ab}^2}=p_{Ab}(1-p_{Ab})+O(p_{Ab}^3).
\end{equation}

Even by neglecting the $O(p_{Ab}^3)$ term, this equation does not have a closed
solution \cite{foot1}.
Yet, it is possible to extract a conclusion concerning stability and  convergence rate. 
Successive iterations of equation (\ref{P01paralambda0v2}) give 
\begin{equation}
p_{Ab}=p_{Ab}^0\sum_{k=0}^{2^t-1} (-1)^k a_k (p_{Ab}^0)^k,
\end{equation}
where $a_0=1$ and $a_1=t$. Accordingly, for times $t< O(1/p_{Ab}^0)$ and points close to
the fixed point along ${\mathbf v_2}$,
\begin{equation}
p_{Ab}\approx p_{Ab}^0(1-t p_{Ab}^0)\approx\frac{p_{Ab}^0}{1+p_{Ab}^0 t},
\end{equation}
which again demonstrates stability, however a  convergence results superslow
when compared with points $E_B$. Numerical computations demonstrate that the
right hand  result is still valid for times arbitrarily large (see appendix
\ref{appendixE}).

\subsection{Conserved quantities}\label{conserved}

Quantities not changing in time give powerful insights in the understanding of
dynamical problems. In the absence of restrictions in reproduction, allele
frequencies $\tilde{p}_A$ and $\tilde{p}_B$ remain constant and this property
characterizes the equilibrium
(\ref{constant_allele_frequencies_1}-\ref{constant_allele_frequencies_4}).
Surprisingly, the dynamics under genetic restrictions   has also a conserved
quantity that, being different from the frequencies of the alleles, allows for
a complete description of the dynamics and the equilibria.

\begin{figure}\centering
\includegraphics[scale=0.6]{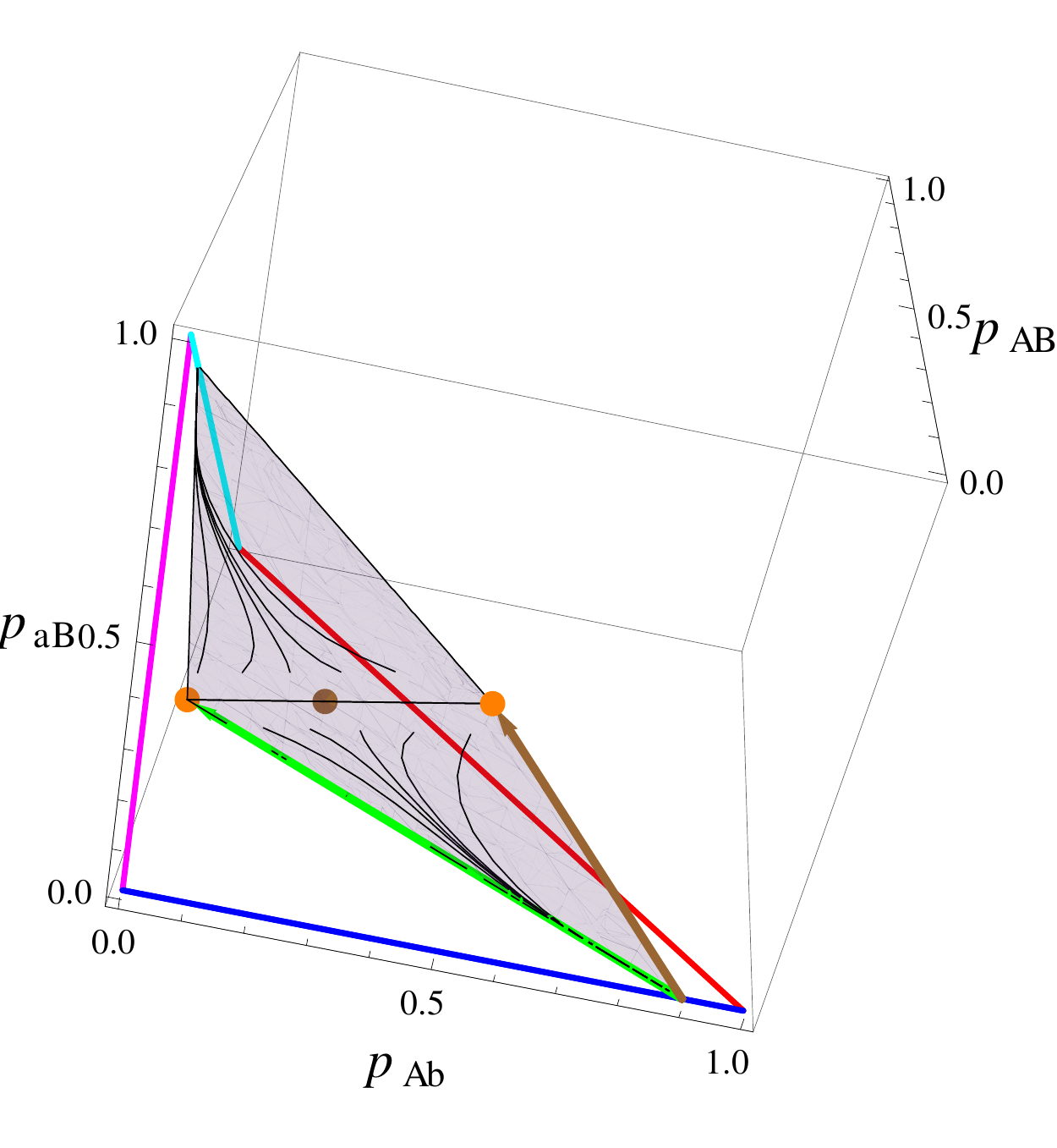}\hfil
\includegraphics[scale=0.6]{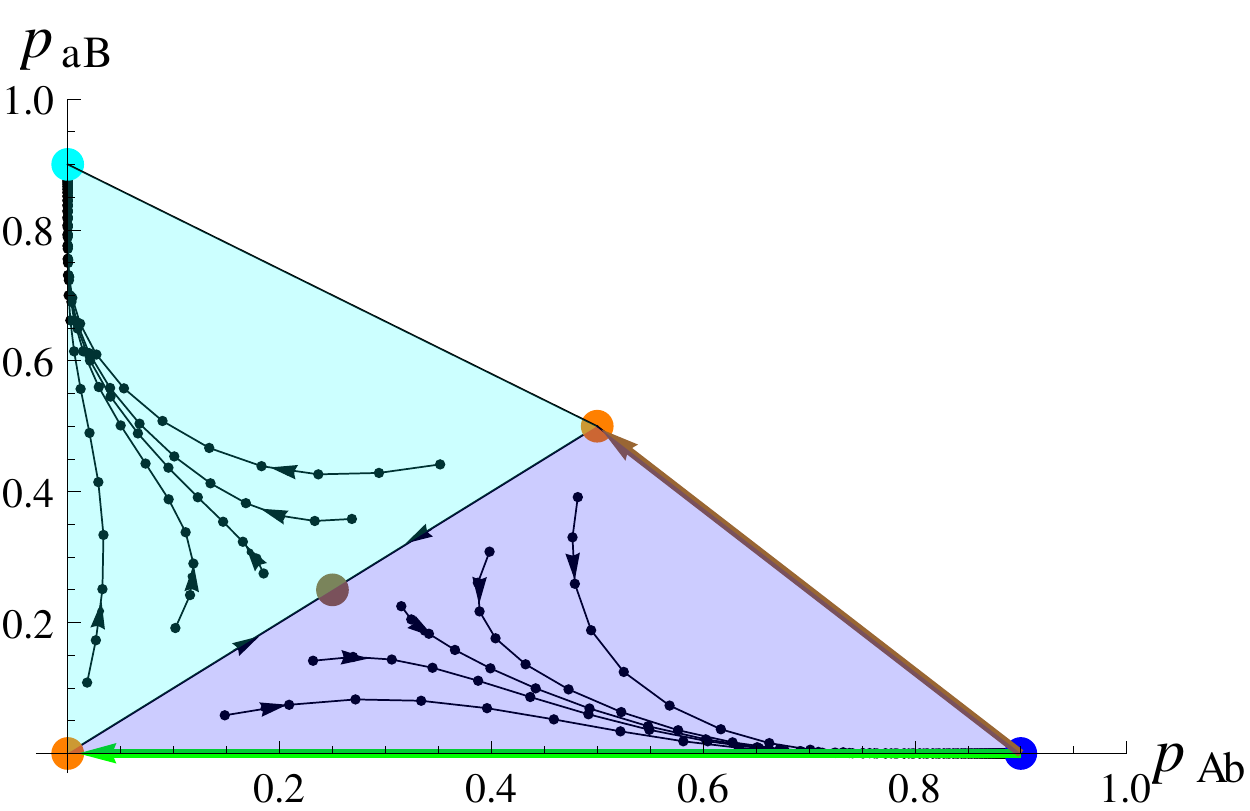}\hfil
\caption{(Color online) Plane of motion corresponding to $T=-0.8$,  and a set of
trajectories with initial conditions chosen close to the stable manifold of
$ES$. In the right pannel, a projection of the T-plane on the
$p_{Ab}\!-\!p_{aB}$ plane. Arrows indicate the direction of motion, and shadowed
regions represent the basins of attaction of fixed points $E_b$ (cyan) and $E_B$
(blue) (compare with figure \ref{ptos_fixos}). In green and brown, the
eigenvectors of the equilibrium $E_B$. Notice the bending of the trajectories
towards ${\bf v_1}$ (green vector), making evident the differential rates of 
convergence in  the two eigendirections. In the picture $\lambda_B=0.9$, which
corresponds to $\tau_1/\tau_2\approx 21.8$.}
\label{trajectories}
\end{figure}

Through equations (\ref{freq_comRes_1}-\ref{freq_comRes_4}) it can be shown 
that all allele frequencies  obey the same evolution equation 
\begin{equation}\label{tilde_menosMedioP}
  \tilde{p}_u^{t+1}=  \frac{\tilde{p}_u-1/2}{1-2\Delta}+1/2
\end{equation}
for $u=A, B, a, b$. Writing this equation for $u=A$ and $u=B$ and
dividing one by the other implies that the quantity
\begin{equation}\label{T}
T=  \frac{\tilde{p}_{A}-1/2}{\tilde{p}_{B}-1/2}
\end{equation}
remains constant across generations. This implies that in the 
3-dimensional  haplotype phase space, the dynamics is constrained to the 
plane defined by the equation
\begin{equation}
p_{AB}+p_{Ab}-1/2-T(p_{AB}+p_{aB}-1/2)=0,
\end{equation}
referred from now on as $T$-plane. Notice that the three aligned points $EU_1$,
$EU_2$ and $ES$ are contained in the $T$-plane for any $T$. Changing the value
of $T$ simply rotates the $T$-plane  around the axis defined by these three
points, making the description of the dynamics quite simple. Specifically,
the location  of the $T$-plane unambiguously determines two stable fixed points,
which can be 
    \begin{enumerate}
     \item $E_B$ and $E_b$ for   $|T|<1$
     \item $E_A$ and $E_a$ for   $|T|>1$
     \item $E_{AB}$ and $E_{ab}$ for $T=1$
     \item $E_{Ab}$ and $E_{aB}$ for $T=-1$.
    \end{enumerate}
The stable eigenvectors ${\bf v}_1$ and ${\bf v}_2$, in turn, run along the borders 
of the  $T$-plane. The dynamics reduces therefore to  a 2-dimensional  
hyperbolic motion with the
central fixed point $ES$ attracting trajectories  in one direction
(corresponding to the $EU_1-ES-EU_2$ axis) and repelling in the other
direction. The latter, unstable direction, gives rise to the unstable manifold
connecting $ES$ with two stable fixed points (in any of the four combinations
listed above). Figure  \ref{trajectories} illustrates the picture  for
$T=-0.8$. 
  
From the previous paragraph results that  by setting the plane of motion,
initial conditions almost determine the  equilibrium distribution of   
haplotype frequencies. There is still an ambiguity concerning which of the two
stable fixed points intersected by the $T$-plane is attained. Of course, this
ambiguity is solved by determining to which side respect to the the
$EU_1-ES-EU_2$ axis the initial condition is located. As we demonstrate next, a
simple algorithm to determine the latter issue consists  in computing initial
values of (\ref{allele_A}-\ref{allele_b}), and identifying the  allele in the
minor proportion.
   
The right panel of Figure \ref{ptos_fixos} depicts a division of the haplotype
space in four regions, and two planes forming the frontiers between  them.
These planes correspond to $T=1$, and $T=-1$. In terms of the alleles
frequencies, a straightforward calculation shows that  on the $(+1)$-plane
$\tilde{p}_A=\tilde{p}_B$, whereas  on the $(-1)$-plane $\tilde p_{A}+\tilde
p_{B}=1$. Accordingly, in one and only one of the four regions, the  alleles
frequencies should satisfy
    \begin{itemize}
     \item $\tilde p_A<\tilde p_B$
     \item $\tilde p_A+ \tilde p_B<1$
    \end{itemize}
but   the second relation implies $\tilde p_A<\tilde p_b$, which necessarily
means $\tilde p_A<1/2$ and thus $\tilde p_A<\tilde p_a$. This  region  of the
haplotype space is therefore characterized by the fact that the allele $A$ is
the allele in the minor proportion.  As  type 1 fixed points labeled $E_A$ (in
purple  in Figure \ref{ptos_fixos}) have necessarily this property, it turns  
out that the region in consideration  must contain all points in the phase 
space that are attracted to this set of fixed points. The conclusion is that
points shadowed in light purple in Figure \ref{ptos_fixos} are the points with
allele A in the minor proportion, and evolve to fixed points $E_A$.  Similar
arguments allow to conclude that the light blue region contains initial
conditions having allele $B$ in the minor proportion (evolving to fixed
points $E_B$), light red region contains initial conditions with allele $a$ in
the minor proportion (evolving to fixed points $E_a$), and finally, light cyan
region contains initial conditions with allele $b$ in the minor proportion
(evolving to fixed points $E_b$).

\subsection{Equilibrium}\label{equilibrium} 

The construction presented above allows us to predict, for an arbitrary initial
condition, the asymptotic equilibrium of the population in terms of two
elements. First, it is necessary to establish the $T$-plane where the initial
condition is located and, second, the allele in the smaller proportion. In the
example of Figure \ref{trajectories}, a bunch of trajectories is simulated
taking initial conditions close to the $EU_1-ES-EU_2$ axis and having $T=-0.8$.
The $T$-plane intersects the $E_B$ set for the initial conditions having $B$ in
the smaller proportion and the $E_b$ set for initial conditions having $b$ in
the smaller proportion. From the conservation of $T=-0.8$ results that at
equilibrium the first bunch of  trajectories converge to the point given by
$\lambda_B=0.9$, and  the second bunch of trajectories to the point given by
$\lambda_b=0.1$. 

The practical result of this analysis is that the smallest among the initial
allelic frequencies always goes to zero. This information, together with the
conserved quantity $T$ suffices to determine all frequencies. For example, if
$p_{b}$ is the smallest initial frequency, in the equilibrium $p_b=0$ and,
consequently, $p_B=1$. From equation (\ref{T}) we find  $p_A = (1+T)/2$ and
$p_a=1-p_A=(1-T)/2$ and all haplotype frequencies have been calculated.

\section{Conclusions and biological implications}\label{conclusions}

The procedure outlined in sections \ref{conserved} and \ref{equilibrium} to
predict the equilibrium from  the initial conditions, in addition to the
information provided in section \ref{rates} concerning times to convergence, 
represent the full solution of the dynamics of the two-loci problem subjected to
genetic restricted mating.  Geometrically, the dynamics reduces to a foliation 
of the 3-dimensional haplotype space in planes with a very  simple motion,
consisting of a central hyperbolic point repelling trajectories towards two
stable equilibria. Initial conditions and stable equilibria remain related 
through the existence of a conserved quantity $T$, which defines the planes
where the  hyperbolic motion takes place. 
   
On the basis of times to convergence, stable equilibria can be divided in two
categories. Stable equilibria of type 1  are attained exponentially, whereas
type 2 equilibria are attained at much slower rates
($\delta^t=\frac{\delta^0}{1+\delta^0 t}$, $\delta$ being the  distance to the
fixed point). It is interesting to observe that this classification has a
biological counterpart. Specifically, slow-attained equilibria  represent
monomorphic populations, whereas exponentially-attained equilibria correspond to
populations that are polymorphic at a single locus. Double-polymorphic
populations are unstable (points of type 3 and 4) or evolve to any of the
former scenarios, reveling the fact that genetic restricted mating has the net
effect of a selection. As pointed out in \cite{gavrilets_dynamical_1999}, models
of incompatibility based on genetic distance have two alternative
interpretations.  One interpretation corresponds to sexual haploid populations
with fitness assigned to pairs of individuals (here fitness is included in the
rate $r_{h_1:h_2}$), and the second interpretation concerns diploid populations
reproducing through random mating, where fitness is a function of 
individual  heterozygosity. Accordingly, the model studied in this work
describes an  evolution process that eliminates double polymorphism  (first
interpretation) or alternatively a selection against double heterozigotes
(second interpretation).  Selection against heterozigote is also known as
underdominance, and explains the disruptive selection causing sympatric
speciation \cite{smith}.
  
Section \ref{conserved} revels another important aspect of restriction through
genetic distance, which concerns the fact that the allele initially appearing in
the smallest proportion remains always in the minor proportion, and vanish 
when the equilibrium is attained. The existence of the conserved quantity $T$,
in turn, also has an interesting consequence from the biological point of view.
As $\tilde p_A=1/2$ represents the maximum polymorphism at the first locus,
$d_A=(\tilde p_A-1/2)^2$ can be interpreted as a measure of the monomorphism for
that gene. Accordingly, the fact that $T^2=d_A/d_B$ remains constant along
evolution,  establishes that the correlation between the polymorphism at the
two loci does not change.
  
In the general situation of $B$ genes and mating genetic restriction by a
distance $G$, stable equilibria are expected to be of $G+1$ different types, in
the form of full monomorphic populations, polymorphic populations  at  a single
locus, polymorphic populations at only two loci, etc, up to polymorphic
populations at the $G$ loci. Accordingly, such scenarios can be related to an
elimination of $G\!+\!1$-uple to $B$-uple polymorphic populations, or
alternatively as a selection against $G\!+\!1$-uple to $B$-uple heterozygotes.
These results will be demonstrated in a subsequent publication. In a spatially
structured population it might happen that different regions converge to
different equilibria, resulting in reproductively isolated species as obtained
in \cite{de_aguiar_global_2009}. As the case studied here exhibits reproductive
isolation only in the trivial way accomplished  by monomorphic species (for
instance, populations  $AB$ and $ab$ are isolated with genetic distance within
the populations $d_w=0$), it is of special interest to explicitly consider the
case $B=3$ and $G=1$. In this situation, populations $ABC\!-\!ABc$ and 
$abC\!-\!abc$ are reproductively isolated with $0<d_w<1$. Moreover, the
existence of a third single polymorphic species $ABc\!-\!Abc$, or even the
monomorphic species $Abc$, may create an ring structure, revealing the richness
of  scenarios that can be realized through this simple arrangement. From the
analysis of the times to convergence of section \ref{rates}, it is also expected
that times to convergence for $G=1$ will behave in the same way  even for $B>2$,
displaying an exponential behavior for single polymorphic species, and a
superslow convergence for monomorphic species. These times to convergence
should eventually be compared with the time to fixation driven by random 
drift. As the model studied here assumes infinite size populations, the model
should be modified to take finite populations into account. A possible way to
estimate the time to fixation driven by random drift would be a Moran
approach \cite{moran,moranaguiar}. The case $B=3$ and $G=2$, on the other hand,
represents also an interesting issue to investigate, as it is expected to
display three different time scales of convergence, corresponding to three types
of stable equilibria.

The scenarios described in the previous paragraph, as well as the influences of
mutations on the results of section \ref{con_restr}, will be the subject of a
future work. Nevertheless, we stress the importance of the study accomplished so
far, as it reveals aspects of the  dynamics that necessarily help to undertake
the analysis  in more complex frameworks. \\

\noindent Acknowledgments. We thank Yaneer Bar-Yam for helpful comments and
discussions. This work was partly supported by FAPESP (Funda\c{c}\~ao de Amparo
\`a Pesquisa do Estado de S\~ao Paulo) and CNPq (Conselho Nacional de
Desenvolvimento Cient\'ifico e Tecnol\'ogico).\\

%%%%%%%%%%%%%%%%%%%%%%%%%%%%%%%%%%%%%%%%%%%%%%%%%%%%%%%%%%%%%%%%%%%%%%%%%%%%%%%%%%%%%%%%%%%%%%%%%%%%%%%%%%%%%%%%%%%%%%%%%
 \begin{appendix}
%%%%%%%%%%%%%%%%%%%%%%%%%%%%%%%%%%%%%%%%%%%%%%%%%%%%%%%%%%%%%%%%%%%%%%%%%%%%%%%%%%%%%%%%%%%%%%%%%%%%%%%%%%%%%%%%%%%%%%%%% 

\section{Evolution of the haplotypic frequencies under random mating}
\label{appendixA}
 
In this appendix we solve equations (\ref{sr11}-\ref{sr14}), performing 
explicit calculations for the expression (\ref{sr11}). The remaining
solutions can be obtained in an equivalent way. We write explicitly the time
dependence  

\begin{equation}\label{ap_1}
  p_{AB}^{t+1} = p_{AB}^t - \frac{1}{2}D^t,
\end{equation}
or
\begin{eqnarray}\label{ap_2}
 p_{AB}^1 &=& p_{AB}^0 - \frac{1}{2}D^0\nonumber \\
 p_{AB}^2 &=& p_{AB}^0 -\frac{1}{2}D^1- \frac{1}{2}D^0\nonumber \\
 p_{AB}^3 &=& p_{AB}^0 -\frac{1}{2}D^2 - \frac{1}{2}D^1 - \frac{1}{2}D^0\nonumber \\
 \vdots \nonumber \\
 p_{AB}^t &=& p_{AB}^0 - \frac{1}{2}\displaystyle\sum_{i=0}^{t-1}D^i.
\end{eqnarray}
On the other hand, using the result (\ref{D_de_t}) yields
\begin{eqnarray}\label{ap_3}
  p_{AB}^t = p_{AB}^0 - \frac{D^0}{2}\displaystyle\sum_{i=0}^{t-1}2^{-i}=p_{AB}^0 - D^0(1 - 2^{-t}),
\end{eqnarray}
For long times, we obtain  the equilibrium solutions, 
\begin{eqnarray}\label{ap_5}
 p_{AB}^{\rm eq} &=& p_{AB}^0 - D^0\nonumber \\
 p_{bB}^{\rm eq} &=& p_{Ab}^0 + D^0\nonumber \\
 p_{aB}^{\rm eq} &=& p_{aB}^0 + D^0\nonumber \\
 p_{ab}^{\rm eq} &=& p_{ab}^0 - D^0.
\end{eqnarray}
 
%%%%%%%%%%%%%%%%%%%%%%%%%%%%%%%%%%%%%%%%%%%%%%%%%%%%%%%%%%%%%%%%%%%%%%%%%%%%%%%%%%%%%%%%%%%%%%%%%%%%%%%%
\section{Relationship between allele and haplotype frequencies}
\label{appendixB}

Here we demonstrate the product relationship between the allele 
and the haplotype frequencies when mating is not restricted by 
genetic distance. Given the definitions of the allele frequencies, we
calculate, for instance, the product $\tilde{p}_{A}\tilde{p}_B$
\begin{eqnarray}\label{apB_1}
\tilde{p}_A\tilde{p}_B &=&(p_{AB} + p_{Ab})(p_{AB} + p_{aB})\nonumber \\ 
&=& p_{AB}(p_{AB} + p_{Ab} + p_{aB}) + p_{Ab}p_{aB}\nonumber\\
&=&p_{AB}(1-p_{ab})+ p_{Ab}p_{aB}\nonumber\\
&=& p_{AB} - D.
\end{eqnarray}

Using the result (\ref{D_de_t}) leads to 
\begin{equation}\label{apB_2}
p_{AB} = \tilde{p}_A \tilde{p}_B + \frac{D^0}{2^t}.
\end{equation}

The equilibrium corresponds to the asymptotic limit of the previous equation,  
and is approached after a small number of generations 
\begin{equation}\label{apB_3}
p_{AB}^{\rm eq} = \tilde{p}_A \tilde{p}_B.
\end{equation}
Strictly  speaking, $|p_{AB} -\tilde{p}_A \tilde{p}_B|<1\%$ of the equilibrium
value $\tilde{p}_A \tilde{p}_B$ in less than ten generations.

%%%%%%%%%%%%%%%%%%%%%%%%%%%%%%%%%%%%%%%%%%%%%%%%%%%%%%%%%%
%%%%%%%%%%%%%%%%%%%%%%%%%%%%%%%%%%%%%%%%%%%%

\section{Stability of fixed points of type 1 and 2 (complement)}\label{appendixC}

In this appendix we extend the results of the stability analysis of section 
\ref{stab_anal} for the fixed points $E_A$, $E_B$ and $E_b$. We start with the
eigenvalues and eigenvectors of the stability matrices of the different fixed
points.

\noindent  {\bf Stability of points $E_A$ (purpple line, $p_{aB}=\lambda_A)$}
\begin{itemize}
 \item $\zeta_1=\lambda_A  $:   $\quad \qquad {\mathbf v_1}=\frac{1}{2}{\bf e}_{Ab}-\lambda_A{\bf e}_{aB}$
 \item $\zeta_2=1-\lambda_A$:   $\quad {\mathbf v_2}=\frac{1}{2}{\bf e}_{Ab}+(\frac{1}{2}-\lambda_a){\bf e}_{aB}$.
 \item $\zeta_3=1          $:   $\quad \qquad {\mathbf v_3}={\bf e}_{aB}$.
 \end{itemize}

\noindent  {\bf Stability of points $E_a$ (red line, $p_{AB}=\lambda_a$, $p_{Ab}=1-\lambda_a$)}
\begin{itemize}
 \item  $\zeta_1=1-\lambda_a$: $\quad {\mathbf v_1}=(\frac{1}{2}-\lambda_a){\bf e}_{AB}+(-1+\lambda_a){\bf e}_{Ab}$.
 \item  $\zeta_2=\lambda_a$:   $\quad \qquad {\mathbf v_2}=-\lambda_a{\bf e}_{AB}+(-\frac{1}{2}+\lambda_a){\bf e}_{Ab}+{\bf e}_{aB}$
 \item $\zeta_3=1$: $\quad \qquad {\mathbf v_3}={\bf e}_{AB}-{\bf e}_{Ab}$.
 \end{itemize}

\noindent  {\bf Stability of points $E_b$ (cyan line, $p_{AB}=\lambda_b$, $p_{aB}=1-\lambda_b$)}
\begin{itemize}
 \item  $\zeta_1=1-\lambda_b$: $\quad {\mathbf v_1}=(\frac{1}{2}-\lambda_b){\bf e}_{AB}-(1-\lambda_b){\bf e}_{aB}$.
 \item  $\zeta_2=\lambda_b$:   $\quad \qquad {\mathbf v_2}=-\lambda_b{\bf e}_{AB}+{\bf e}_{Ab}+(-\frac{1}{2}+\lambda_b){\bf e}_{aB}$
 \item $\zeta_3=1$: $\quad \qquad {\mathbf v_3}=-{\bf e}_{AB}+{\bf e}_{aB}$.
 \end{itemize}

\begin{figure}
\centering
\begin{subfigure}{.49\textwidth}
  \centering
  \includegraphics[scale=0.49]{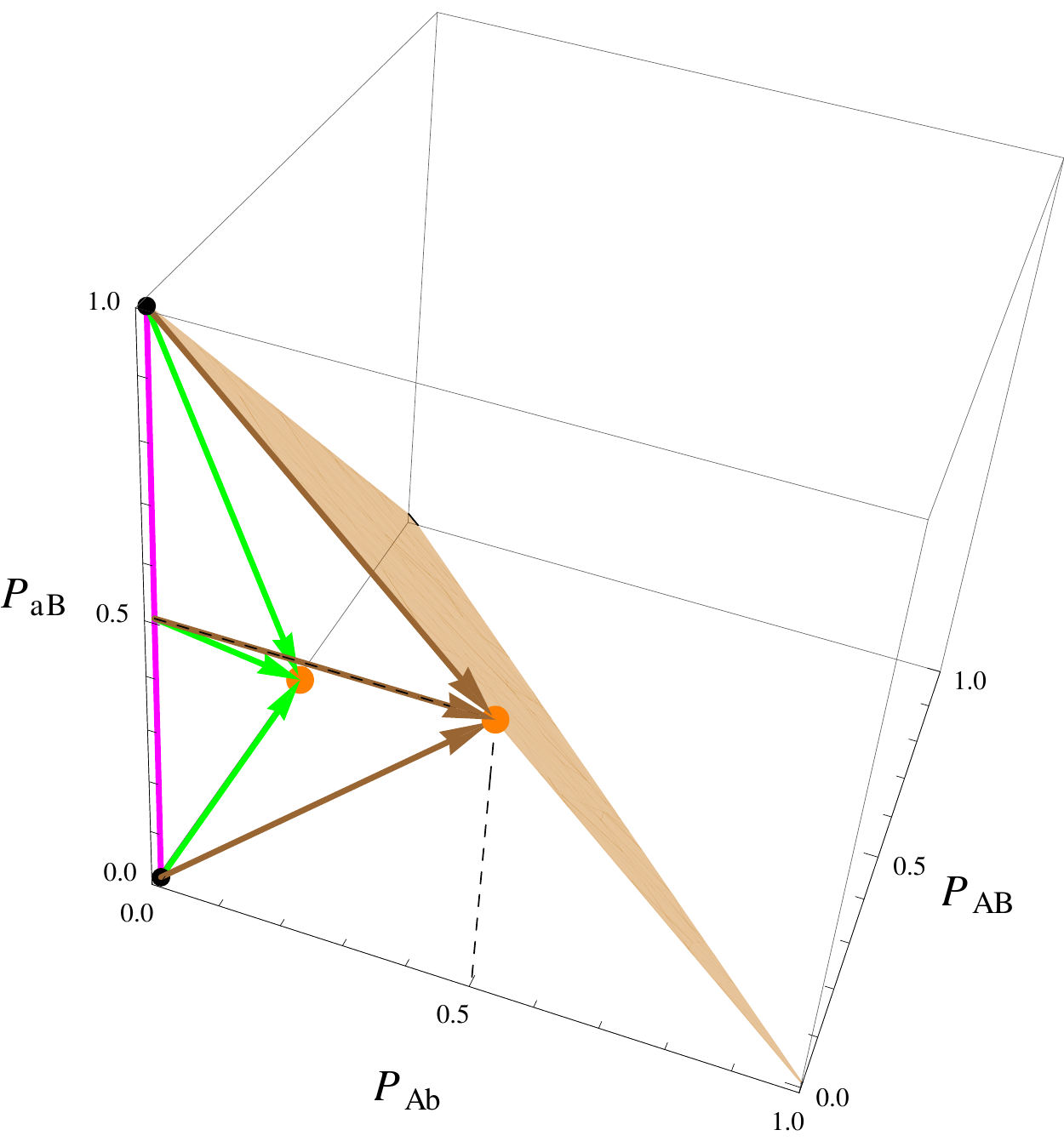}
 % \caption{Eigenvector for fixed points p1a}
  \label{p1a}
\end{subfigure}%
\begin{subfigure}{.49\textwidth}
  \centering
  \includegraphics[scale=0.49]{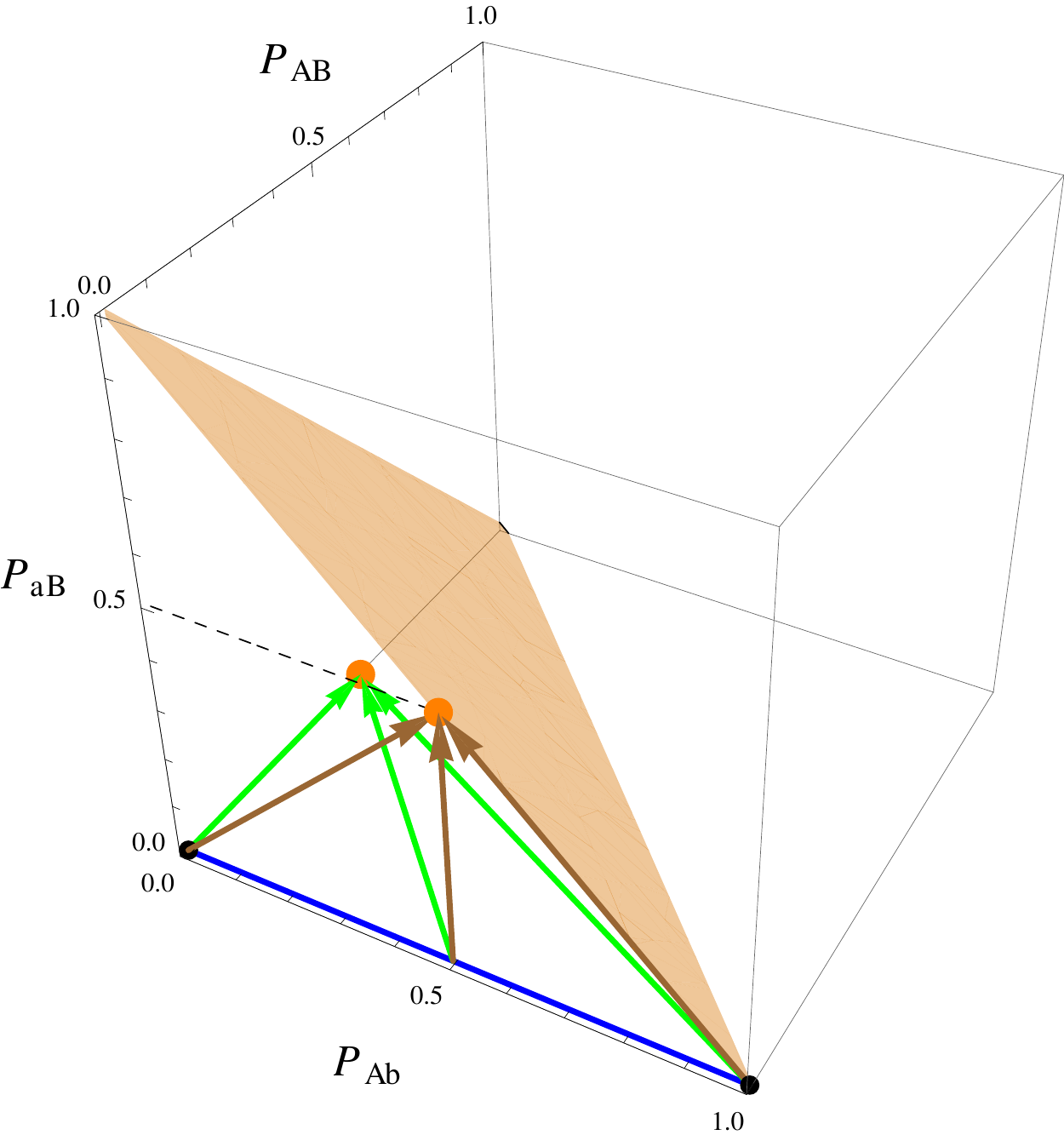}
  %\caption{Eigenvector for fixed points p1c}
    \label{p1c}
\end{subfigure}
\begin{subfigure}{.49\textwidth}
  \centering
  \includegraphics[scale=0.49]{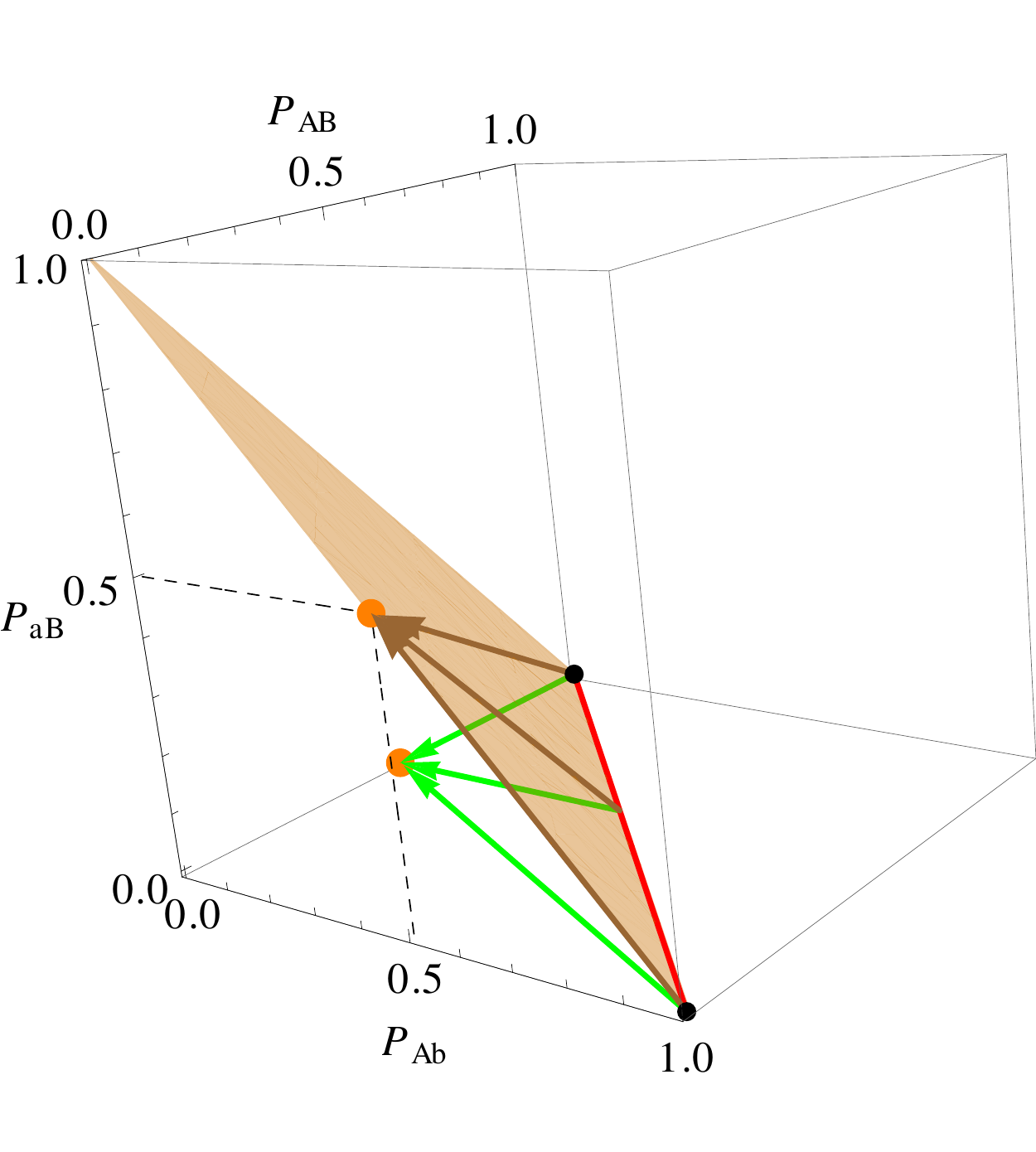}
  %\caption{Eigenvector for fixed points p1d}
    \label{p1d}
\end{subfigure}
\begin{subfigure}{.49\textwidth}
  \centering
  \includegraphics[scale=0.49]{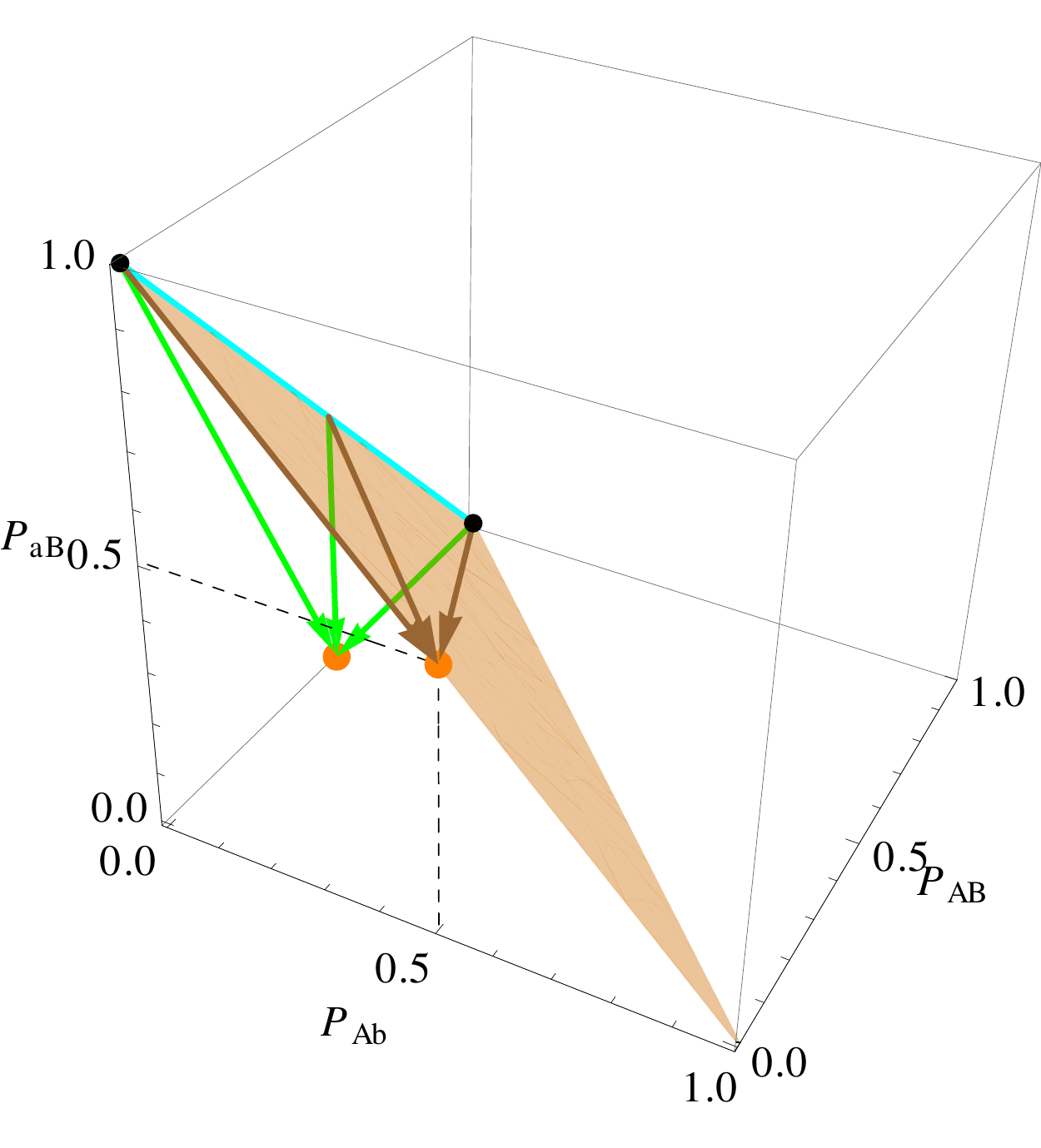}
  %\caption{Eigenvector for fixed points p1d}
    \label{p1d}
\end{subfigure}
\caption{(Color online) Eigenvectors corresponding to type 1 and type 2 fixed
points (vectors ${\mathbf v_1}$ in green, and ${\mathbf v_2}$  in brown).}
\label{vectorsitos_1a-c}
\end{figure}

Stable eigenvectors corresponding to points of type 1 and type 2, properly
scaled, connect the fixed points with type 3 fixed points. In Figure
\ref{vectorsitos_1a-c} we expose this important property for some specific 
points at each set, including the points $E_B$ analyzed in section \ref{stab_anal}.

\begin{figure}
\centering
\includegraphics[scale=0.6]{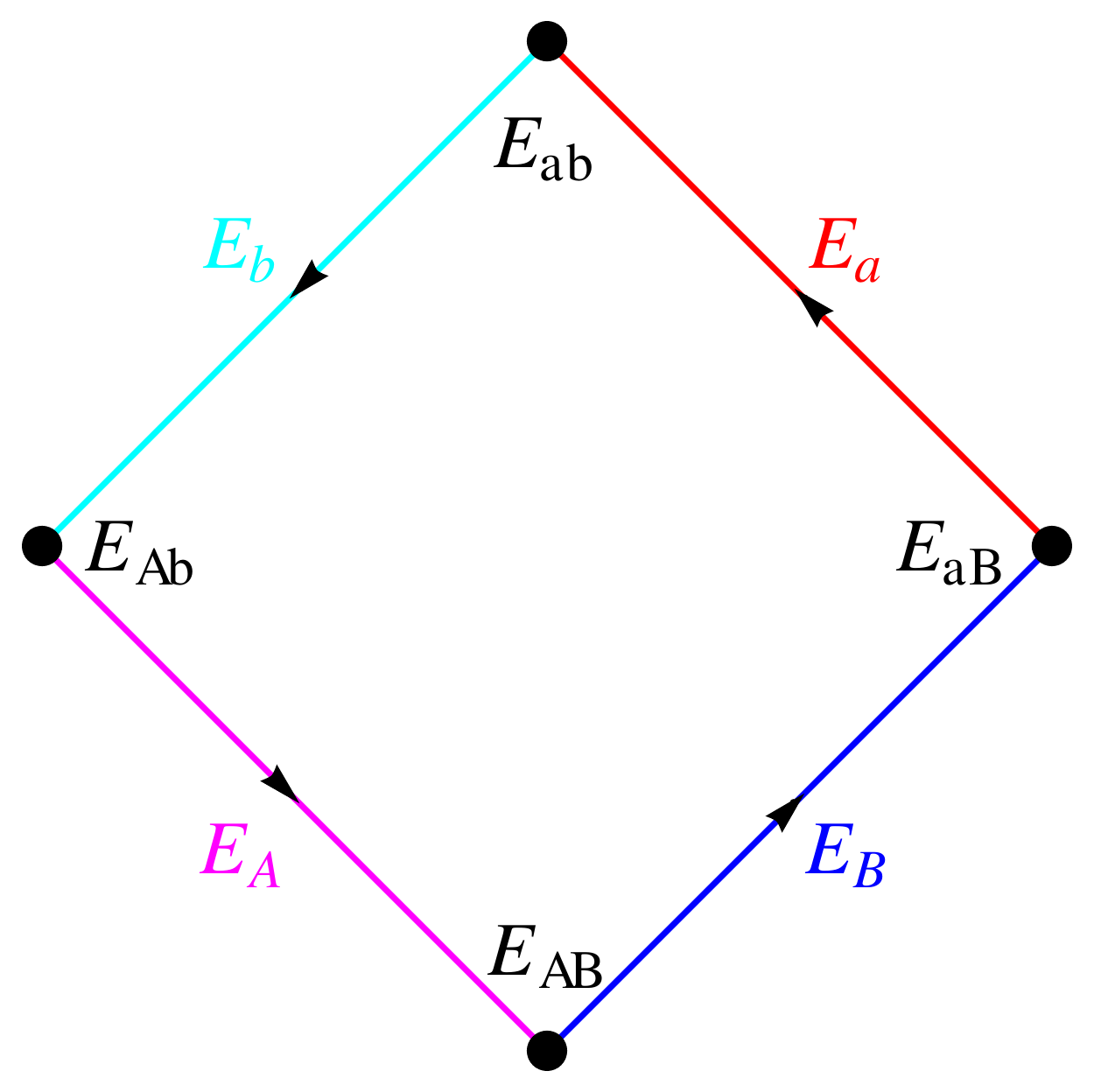}
\caption{(Color online) Schematic representation of type 1 and type 2 fixed
points. Arrows indicate the direction of the walk employed 
in the text to describe the behavior of the rates of convergence.\label{recorrido_bordes_estables_del_Tetrahedro}}
\end{figure}

Rates of convergence are inferred from the relation
\begin{equation}
\tau_i= -\frac{1}{\ln{\zeta_i}}
\end{equation}
where $i=1,2$ denotes the type 3 fixed point ($EU_1$ or $EU_2$) to which the
eigenvector  points. Accordingly, going through the set $E_a$ from $E_{aB}$ to 
$E_{ab}$  (see Figure \ref{recorrido_bordes_estables_del_Tetrahedro}), makes 
$\tau_1$ to decreases. This time becomes almost zero at the point $E_{ab}$
(superfast convergence), and it starts increasing again by going to $E_{Ab}$
through the set $E_b$ (cyan line in Figure
\ref{recorrido_bordes_estables_del_Tetrahedro}). At this $\tau_1$ becomes
infinite (superslow convergence), which means an equivalent  behavior to that
corresponding to  $E_{Ab}$. On the other hand, as $\zeta_2=1-\zeta_1$, it turns
out that $\tau_2$, the time to convergence along the directions spanned by
${\mathbf{v_2}}$, displays the opposite behavior. Finally, the picture is
completed by observing that the  remaining branch of the cycle ($E_{Ab}
\rightarrow E_{AB}$ along $E_A$  and $E_{AB} \rightarrow E_{aB}$  along $E_B$)
is an exact repetition of the branch described above. 

%%%%%%%%%%%%%%%%%%%%%%%%%%%%%%%%%%%%%%%%%%%%%%%%%%%%%%%%%%%%%%%%%%%%%%%%%%%%%%%%%%%%%%%%%%%%%%%%%%%%%%%%

\section{Numerical solution of equation (\ref{P01paralambda0v2})}\label{appendixE}

\begin{figure}
\centering
\begin{subfigure}{.49\textwidth}
  \centering
  \includegraphics[scale=0.5]{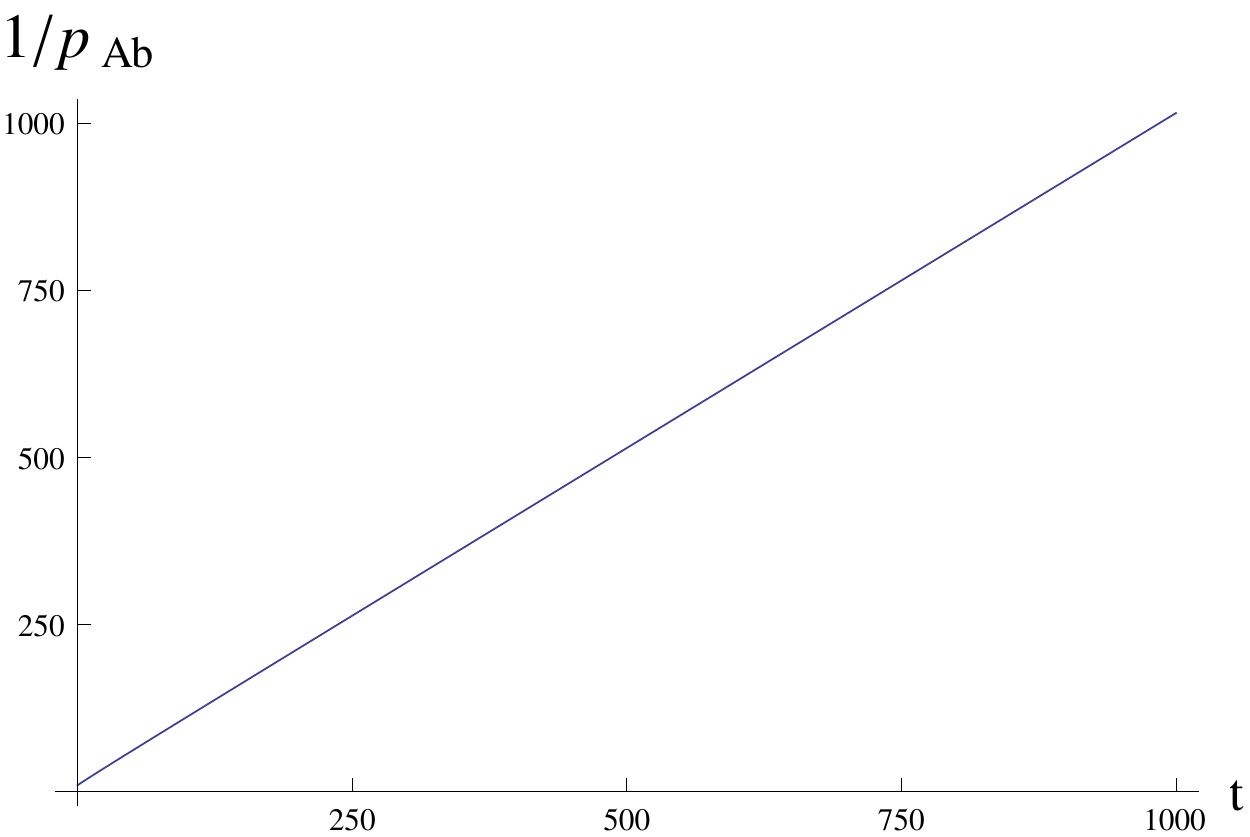}
 \caption{ \footnotesize{(Color online) Iterations of the map for the initial
condition  \mbox{$p_{Ab}^0=0.1$}. The fitting gives
  \mbox{$1/p_{Ab}=12.336+1.002 t $}}.}
  \label{rectaP01}
\end{subfigure}
\begin{subfigure}{.49\textwidth}
  \centering
  \includegraphics[scale=0.5]{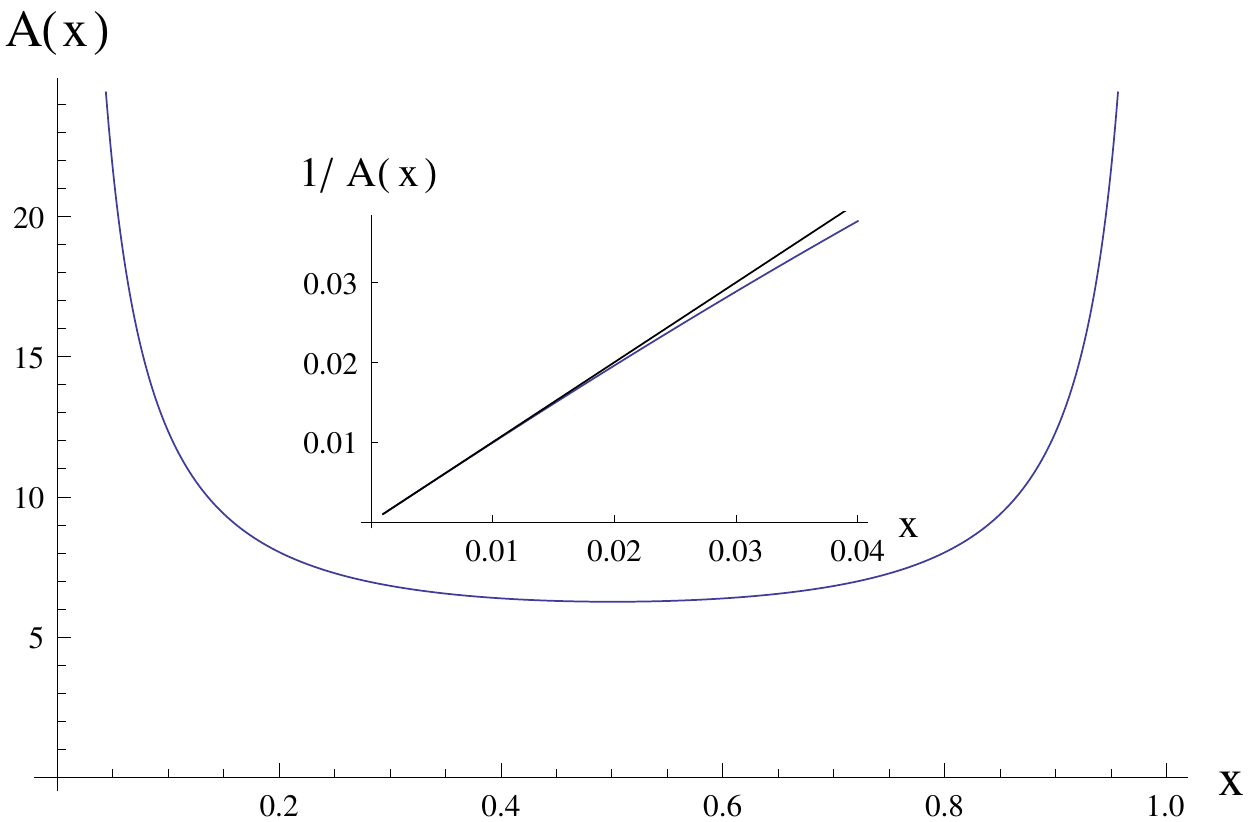}
 \caption{\footnotesize{(Color online) Plot of the function
 $A(x)$ (see the text). In the inset we plot $1/A(x)$  to demonstrate that, for very 
 small $x$ values $1/A(x)=x+O(x^2)$ (black curve: $f(x)=x)$.} }
  \label{rectaP01}
\end{subfigure}
\caption{Fitting results related to the eigenvector {$\mathbf v_2$} of the stability 
matrix associated to $E_{AB}$.  }
\label{adedelta_y_recta}
\end{figure}

Here we give a brief summary of the fitting process employed to solve equation 
(\ref{P01paralambda0v2}). By iterating the map for different initial conditions
and fitting the results, one obtains
\begin{equation}
 p_{Ab}^t=\frac{1}{A(p_{Ab}^0)+t},
\end{equation}
where $A(x)$ is a function  that   diverges as $1/x$ when $x\rightarrow 0$ (see
Figure \ref{adedelta_y_recta}). Accordingly, for very small $p_{Ab}^0$ values we
can write
\begin{equation}
 p_{Ab}^t=\frac{p_{Ab}^0}{1+p_{Ab}^0 t}+O((p_{Ab}^0)^2).
\end{equation}

%%%%%%%%%%%%%%%%%%%%%%%%%%%%%%%%%%%%%%%%%%%%%%%%%%%%%%%%%%%%%%%%%%%%%%%%%%%%%%%%%%%%%%%

\end{appendix}
  

\begin{thebibliography}{32}
\bibitem{foot1} The equation is a particular case of the
logistic map $x_{n+1}=r x_n(1-x_n)$, having analytical
solutions only for $r=-2$, $r=2$ and $r=4$.
\end{thebibliography}
\end{document}